\documentclass[12pt,english,cite]{article}
\usepackage{epsfig}
\usepackage{psfrag}
\title{Neutrinos from SN1987A, Earth matter effects and the LMA solution of the solar neutrino problem}

\author{C.Lunardini$^{a)}$, A.Yu.Smirnov$^{b)}$}

\begin{document}

\def \lta {\mathrel{\vcenter{\hbox{$<$}\nointerlineskip\hbox{$\sim$}}}}
\def \gta {\mathrel{\vcenter{\hbox{$>$}\nointerlineskip\hbox{$\sim$}}}}

\bibliographystyle{unsrt}

\maketitle

\noindent
\begin{center}
{\small {\it a) SISSA-ISAS, via Beirut 2-4, 34100 Trieste, Italy} \\
{\it and INFN, sezione di Trieste, via Valerio 2, 34127 Trieste, Italy}\\
\vspace{0.4cm}
{\it b) The Abdus Salam ICTP, Strada Costiera 11, 34100 Trieste,
Italy} \\
{\it and Institute for Nuclear Research, RAS, Moscow, Russia}}
\end{center}
\vspace{0.5cm}

\begin{abstract}
\noindent
We study properties of the oscillation effects in the matter of the Earth on antineutrino fluxes from supernovae.
We show that these effects can provide explanation of the  difference in the energy spectra of the events detected by Kamiokande-2 and IMB detectors from SN1987A as well as the absence of high-energy events with $E\gta 40$ MeV. This explanation requires the neutrino oscillation parameters $\Delta m^2$ and $\sin^2 2\theta $ to be in the region of the LMA solution of the solar neutrino problem and the normal mass hierarchy if $|U_{e 3}|^2\gta 10^{-3}$.  The hierarchy can be inverted if $|U_{e 3}|^2\ll 10^{-3}$. The solution of the solar neutrino problem based on $\nu_e$-conversion to a pure sterile state is disfavoured by SN1987A data. 
\end{abstract}

%\pacs{PACS: 14.60.Pq, 97.60.Bw}

\newpage

%useful commands:
%{}
%\left{ \right}
%\cite{ }

%\begin{eqnarray} 

%\label{ }
%\end{eqnarray}

%(\ref)

%[]
%\left[ \right]

%%%%%%%%%%%%%%%%%%%%%%%%%%%%%%%%%%%%%%%%%%%%%%%%%%%%%%%%%%%%%%%%%%
%%%%%%%%%%%%%%%%%%%%%%%%%%%%%%%%%%%%%%%%%%%%%%%%%%%%%%%%%%%%%%%%%%%
\section{Introduction}
\label{sect:1}
The detection of the neutrino burst from the supernova SN1987A by the Kamiokande-2 \cite{Hirata:1987h} and IMB \cite{Bionta:1987qt} detectors has confirmed the general picture of gravitational collapse, hot neutron star formation and neutrino emission (see \cite{Suzuki,Raffelt:1996wa} for a review).

At the same time certain features of the detected neutrino signals remain unexplained. 
In this paper we will consider the difference in the energy spectra of the events detected by Kamiokande-2 (K2) and IMB. It is difficult to explain such a difference  in terms of the characteristics of the detectors (energy thresholds, efficiencies of detection, sizes, etc.).  The $\bar{\nu}_e$ spectrum inferred from the K2 events is substantially softer than that from 
the IMB events.
This can be quantified by comparing the parameters of the original $\bar{\nu}_e$ spectra, such as the effective temperature $T_{\bar{e}}$ and the luminosity $L_{\bar{e}}$, that were estimated from the results of the two experiments.  It was found that the regions in the $T_{\bar{e}}-L_{\bar{e}}$ plane determined from the K2 and IMB data have only marginal overlap and the probability that the two sets of data correspond to the same spectrum is less than few per cents \cite{analyses,Janka}. 

More specifically, we will discuss the following features of the SN1987A signals: \\

(i) Concentration of the IMB events in the energy interval 
$E\simeq 35\div 40 $ MeV. \\  
  
(ii) 
Absence of events at IMB above $E\simeq 40$ MeV (which looks like a sharp cut of the 

spectrum). \\

(iii)
Absence of events with $E\gta 35$ MeV at K2.\\

The effects of mixing on supernova neutrinos have been extensively studied
\cite{Mikheev:1986if,mixing}. 
In particular, soon after the observation of SN1987A it was marked that the differences in the K2 and IMB spectra could be related to oscillations of $\bar{\nu}_e$ in the matter of the Earth and to the different positions of the detectors at the time of detection \cite{talkalexei,general}. 
It was realized that, for this oscillations mechanism to work, one needs $\Delta m^2\sim 10^{-5}~{\rm eV^2}$ (i.e. in the region of the Earth regeneration effect) and large (close to maximal) mixing of $\bar{\nu}_e$.

At that time the idea did not attract much attention, since a large lepton mixing was considered unnatural and  conversion with small mixing angle was indicated as the most plausible solution of the solar neutrino problem.   
Moreover, later it was argued \cite{Wolfenstein:1987pj,Smirnov:1994ku} that a large mixing of electron neutrinos is disfavoured by the SN1987A data: the $\bar{\nu}_e \leftrightarrow \bar{\nu}_\mu/\bar{\nu}_\tau$ conversion inside the star leads to the appearance of a high energy tail in the spectrum which contradicts observations.  The bound is absent, however, in the range of Earth matter effects, $\Delta m^2\sim 10^{-5}~{\rm eV^2}$.

Later, the likelihood analysis of the combined K2 and IMB spectra in \cite{Jegerlehner:1996kx} has confirmed the results of \cite{Smirnov:1994ku}. In general, the flavour conversion of $\bar{\nu}_e$ leads to worser agreement with the supernova theory.  Some mild improvement of the likelihood appears for oscillation parameters in the Large Mixing Angle (LMA) region when the Earth matter 
effect is taken into account.  This improvement is  obtained, however, at the price of aggravating the conflict with the SN theory: it was found that the best fit point corresponds to too low temperature of the original $\bar{\nu}_e$  spectrum, 
$T_{\bar{e}}\simeq 1.9$ MeV, and too high total binding energy, $L\simeq 10^{54}$ ergs.  Moreover, the best fit values $\Delta m^2\simeq (3\div 5) \cdot  10^{-6}~{\rm eV^2}$, found  in \cite{Jegerlehner:1996kx} are now excluded by the Superkamiokande data on the day-night asymmetry. 
\\

Since the first proposal \cite{talkalexei} the situation has drastically changed. With very high confidence level the atmospheric neutrino data are explained by 
 $\nu_\mu \leftrightarrow \nu_\tau$ oscillations with maximal or nearly maximal mixing \cite{Fukuda:1998mi}.  The LMA  solution gives the best global fit of all the available solar neutrino data \cite{Gonzalez-Garcia:2000sk}. It seems that large mixing is a general property of leptons.  
Clearly,  if the LMA is the true solution of the solar neutrino problem, then a significant part of the $\bar{\nu}_e$ events detected from SN1987A were produced by the converted  muon and tau antineutrinos. This means that in 1987 we observed the first appearance signal of neutrino conversion!

In view of this we have reconsidered the explanation of the features of the neutrino signals from SN1987A in terms of neutrino conversion in the star and in the Earth. 

In section \ref{sect:2} we study the general properties of the conversion of antineutrinos in the matter of the supernova and of the Earth.  We apply the results to the explanation of the difference of signals detected by Kamiokande-2 and IMB in section \ref{sect:3}.  In section \ref{sect:4} we summarize implications of the Earth matter effect on antineutrinos for the neutrino mass spectrum.
      
\section{Neutrino conversion in the star and in the Earth}
\label{sect:2}
We assume that the electron neutrino mixes with some combination of the muon and tau neutrinos, $\nu_\mu'$, ($\nu_e=\cos\theta \nu_1+\sin\theta \nu_2 $, $\nu_\mu'=\cos\theta \nu_2-\sin\theta \nu_1 $) and the oscillation parameters of the $\nu_e-\nu_\mu'$ system are in the region of the large mixing angle solution of the solar neutrino problem:
\begin{eqnarray} 
\sin^2 2\theta\simeq 0.6\div 1.0~, ~~~~~ \Delta m^2 \simeq (1.5 \div 10)\cdot 10^{-5}~{\rm eV^2}~,
\label{eq:lma}
\end{eqnarray}
(the case of the LOW solution will be considered later).  We will also assume that other $\Delta m^2$ and mixings in the neutrino spectrum are irrelevant for the antineutrino channel.
 
With the parameters (\ref{eq:lma}) the propagation of antineutrinos inside the star is completely adiabatic, so that the following transitions occur \cite{Dighe:2000bi}:
\begin{eqnarray} 
\bar{\nu}_e \rightarrow \bar{\nu}_1~,~~~~ 
\bar{\nu}_\mu' \rightarrow \bar{\nu}_2~.
\label{eq:trans}
\end{eqnarray}
That is, the originally produced electron antineutrino will reach the surface of the star, and consequently the surface of the Earth, as the pure mass eigenstate $\bar{\nu}_1$, and $\bar{\nu}_\mu'$ as the pure mass eigenstate $\bar{\nu}_2$.  As a result, the 
flux of the electron antineutrinos at the surface of the Earth equals:
\begin{eqnarray} 
F_{\bar{e}}=\cos^2 \theta F^0_{\bar{e}}+\sin^2 \theta F^0_{\bar{\mu}'}
= F^0_{\bar{e}} - \sin^2 \theta (F^0_{\bar{e}}- F^0_{\bar{\mu}'})~,
\label{eq:flux}
\end{eqnarray}
where $F^0_{\bar{e}}$ and $F^0_{\bar{\mu}'}$ are the original fluxes of $\bar{\nu}_e$ and $\bar{\nu}_\mu'$ produced in the center of the star.

In the matter of the Earth the mass eigenstates $\bar{\nu}_1$ and $\bar{\nu}_2$
oscillate.  Taking into account these oscillations we find the flux of $\bar{\nu}_e$ at the detector \cite{Dighe:2000bi}:
\begin{eqnarray} 
F^D_{\bar{e}}
= F_{\bar{e}} + (P_{1 \bar{e}}- \cos^2 \theta) (F^0_{\bar{e}}- F^0_{\bar{\mu}'})~,
\label{eq:fluxd}
\end{eqnarray}
where $P_{1 \bar{e}}$ is the $\bar{\nu}_1 \rightarrow \bar{\nu}_e$ transition probability in the matter of the Earth; in the derivation of eq. (\ref{eq:fluxd}) we used also the relation $P_{2 \bar{e}}=1-P_{1 \bar{e}}$.
Since for both K2 and IMB the neutrino trajectories were in the mantle, where the density change is rather small, we use for $P_{1 \bar{e}}$ the formula for oscillations in uniform medium \cite{Smirnov:1994ku}:
\begin{eqnarray} 
P_{1 \bar{e}}=\cos^2 \theta + \sin 2\theta_m \sin\left(2\theta - 2\theta_m \right) 
\sin^2 \left({\pi d \over l_m} \right)~.
\label{eq:unif}
\end{eqnarray}
Here $d$ is the length of the neutrino trajectory inside the Earth for a given detector, $\theta_m$ and $l_m$ are the mixing angle and the oscillation length for the antineutrinos in the matter of the Earth:
\begin{eqnarray} 
\sin^2 2\theta_m={\sin^2 2\theta \over {(\cos 2\theta + (l_\nu/ l_0))^2+ \sin^2 2\theta}}~,~~~~~~~~~~ l_m=l_\nu { \sin 2\theta_m\over \sin 2\theta}~.
\label{lmix}
\end{eqnarray}
In eq. (\ref{lmix}) $l_{\nu}\equiv 4\pi E/\Delta m^2$ and 
$l_0\equiv 2\pi/(\sqrt{2} G_F n_e)$ are the vacuum oscillation length  and 
the refraction length; $G_F$ is the Fermi coupling constant and $n_e$ is the electron number density. In the expression (\ref{eq:unif}) $\theta_m$ and $l_m$ should be taken for the average electron number density $n_e$ along the 
neutrino trajectory for a given detector. \\

The qualitative features of the conversion effects can be immediately seen from eqs. (\ref{eq:flux})-(\ref{eq:unif}), taking into account that the original $\bar{\nu}_\mu'$  flux has harder spectrum than the one of $\bar{\nu}_e$ and the total luminosities in the different flavours are comparable.
This means that a critical energy, $E_c$, exists such that:
\begin{eqnarray} 
&&F^0_{\bar{\mu}'}> F^0_{\bar{e}}~~~{\rm for}~~~ E>E_c \nonumber \\
&&F^0_{\bar{\mu}'}< F^0_{\bar{e}}~~~{\rm for}~~~ E<E_c~. 
\label{eq:ec}
\end{eqnarray}
The energy $E_c$   depends, in particular, on the effective temperatures  and luminosities of the original $\bar{\nu}_\mu'$ and  $\bar{\nu}_e$ fluxes:
$E_c=E_c(T_{\bar{e}},T_{\bar{\mu}'},L_{\bar{e}},L_{\bar{\mu}'})$.

According to eq. (\ref{eq:flux}) the conversion in the star leads to a composite spectrum of electron antineutrinos with $\sin^2 \theta$ admixture of the hard component which generates an high energy tail in the sample of events.
In comparison with the original  $\bar{\nu}_e$ spectrum, $F^0_{\bar{e}}$, the flux  $F_{\bar{e}}$  is suppressed at 
$E<E_c$  and enhanced  at  $E>E_c$. The effect is proportional to $\sin^2 \theta$ and the conversion probability itself does not depend on the neutrino energy.

The oscillations inside the Earth give an opposite effect   
(see eqs. (\ref{eq:fluxd}), (\ref{eq:unif})): they enhance the flux at $E<E_c$
 and suppress it at high energies. Notice that, since $P_{1 \bar{e}}>\cos^2 \theta$, for LMA parameters and in constant density approximation the matter of the Earth always regenerates the $\bar{\nu}_e$: the oscillatory factor in eq. (\ref{eq:unif}) is positive since $2 \theta_m=(0\div \pi)$ and 
$\theta>\theta_m$. 
Shortly, an increase of the medium density will tend to return the neutrino state into the initial state, that is, $\bar{\nu}_e$ at very high densities. 
Entering the Earth is equivalent to such an increase.

Combining eqs. (\ref{eq:flux}) and (\ref{eq:fluxd}) we find:
\begin{eqnarray} 
F^D_{\bar{e}}=F^0_{\bar{e}}- (F^0_{\bar{e}}-F^0_{\bar{\mu}'})(1-P_{1 \bar{e}})~,
\label{eq:combin}
\end{eqnarray}
which shows that the overall conversion effect, both in the star and in the Earth, leads to suppression of the flux at low energies ($E<E_c$)  and to its enhancement at high energies ($E>E_c$). The effect is proportional to the difference of the original fluxes.  \\

Let us consider the properties of the Earth regeneration effect in some detail. The formula (\ref{eq:combin}) can be written as: 
\begin{eqnarray} 
F^D_{\bar{e}}=F^0_{\bar{e}}- (F^0_{\bar{e}}-F^0_{\bar{\mu}'})(\sin^2 \theta -A_p \sin^2 \phi)~, 
\label{eq:combinrew}
\end{eqnarray}
where
\begin{eqnarray} 
A_p\equiv \sin 2\theta_m \sin\left(2\theta - 2\theta_m \right)~,~~~~~
\phi\equiv{\pi d \over l_m}
\label{eq:ap}
\end{eqnarray}
are the depth and the phase of oscillations.
In contrast with the conversion in the star, the Earth matter effect has strong energy dependence.  
The depth of oscillations, $A_p$, equals zero at very low energies, where $\theta_m\simeq \theta$, and at high energies, where $\theta_m\simeq \pi/2$.  The depth reaches the maximum
\begin{eqnarray} 
A_p^{max}=\sin^2 \theta
\label{eq:amax}
\end{eqnarray}
 at $\theta_m=\theta/2$, which corresponds to the equality
\begin{eqnarray} 
{l_{\nu} \over l_0}=1~.
\label{eq:ll0}
\end{eqnarray}
Notice that the depth of oscillations increases with $\theta$.
At the condition  of maximal depth, eq. (\ref{eq:ll0}),  also the conversion probability $P_{1 \bar{e}}$ reaches the maximum: 
\begin{eqnarray} 
P_{1 \bar{e}}=1 ~,
\label{eq:pmax}
\end{eqnarray}
provided that $\phi=\pi/2+k \pi$ with $k=0,1,2,\ldots $~. 
For $E\rightarrow 0$ or $E\rightarrow \infty$, we have $P_{1 \bar{e}}\rightarrow \cos^2 \theta$. Thus,  $A_p$ as well as $P_{1 \bar{e}}$ have a resonance character with eq. (\ref{eq:ll0}) being the resonance condition.  Notice that the condition  (\ref{eq:ll0}) does not depend on 
the mixing angle, in contrast with the resonance condition for the pure flavour case.   

From eq. (\ref{eq:ll0}) we find the resonance energy
\begin{eqnarray} 
E_R={\Delta m^2\over {2 \sqrt{2} G_F n_e}}~.
\label{eq:eres}
\end{eqnarray}
The interval of energies $E_{-}\div E_{+}$  around (\ref{eq:eres}) for which $A_p>A_p^{max}/2=\sin^2 \theta/2$ is determined by: 
\begin{eqnarray} 
{E_{\pm} \over E_R}=\cos 2\theta +2 \pm \sqrt{(\cos 2\theta +1)(\cos 2\theta +3)}~.
\label{eq:width}
\end{eqnarray}
Thus $E_R$  and $E_{\pm}$ give the interval of strong oscillation effect in the Earth. 
%Taking the matter density $\rho\simeq 4~{\rm g\cdot cm ^{-3}}$ and 
%$\Delta m^2\simeq 2\cdot 10^{-5}~{\rm eV^2}$ we find $E_R\simeq 90$ MeV, 
%and for $\cos 2\theta\simeq 0.5$ one gets $E_{-}\simeq 20$ MeV.  
%Thus, a strong Earth matter effect is expected in the most energetic 
%part of the neutrino spectra in K2 and IMB ($E\simeq 20\div 40$ MeV).
 
The peak in the $P_{1 \bar{e}}$ probability exists in the antineutrino channel even if the usual flavour resonance is in the neutrino channel. 
With increase of $\theta$ and its shift to the dark side, 
$\theta \gta \pi/4$,
 the peak gets narrower and at $\theta \rightarrow \pi/2$ it converges to the usual flavour resonance peak.

The oscillation length  $l_m$ decreases with $E/\Delta m^2$; it also decreases with $\cos 2\theta$.

At high energies, due to the exponential decrease of the fluxes with $E$ and 
the difference of temperatures of the $\bar{\nu}_e$ and $\bar{\nu}_\mu'$ 
spectra, we get $F^0_{\bar{\mu}'}\gg F^0_{\bar{e}}$, and therefore, according to eq. (\ref{eq:combin}), the flux at the detector equals:
\begin{eqnarray} 
F^D_{\bar{e}}\simeq F^0_{\bar{\mu}'} (1-P_{1 \bar{e}})~.
\label{eq:fluxapp}
\end{eqnarray}
Then, at the resonance, $E\simeq E_R$, eq. (\ref{eq:eres}),  the $\bar{\nu}_e$ flux can be suppressed completely: $F^D_{\bar{e}}\simeq 0$, if the phase of oscillations equals a semi-integer fraction of $\pi$. This suppression does not depend on the value of 
the mixing angle $\theta$. In what follows we will show that this suppression could indeed be realized in the case of neutrinos from SN1987A.

\section{Neutrino signals in K2 and IMB }
\label{sect:3}
Let us consider the difference of signals at K2 and IMB which can be produced by oscillations in the Earth.  This difference is related to the  distances travelled by the neutrinos in the Earth: $d_{IMB}\simeq 8535$ Km for IMB, $d_{K2}\simeq 4363$ Km for K2, and to the average densities $\rho_{IMB}\simeq 4.5 ~{\rm g\cdot cm^{-3}}$, $\rho_{K2}\simeq 3.5 ~{\rm g\cdot cm^{-3}}$ along the trajectories. As a consequence, both the depths and the phases of oscillations at K2 and IMB are different.

Let us show  that the features (i)-(iii) indicated in the introduction  can be explained by  oscillations in the matter of the Earth. Moreover,  the explanation implies certain values of $\Delta m^2$ and $\sin^2 2\theta$. 
%It should be stressed that, due to the small number of the detected events all% the features mentioned here can be due just to statistical fluctuations and o%ur conclusions have mostly indicative character.

To reproduce the characteristics described in (i)-(iii) we require that:\\

\noindent 
(1) The phase of oscillations at IMB detector at $E\simeq 38\div 42$ MeV equals
\begin{eqnarray} 
\phi_{IMB}(40)\equiv{\pi d_{IMB}\over l_m}=k \pi~,~~~~k=1,2,3,...~~.
\label{eq:phasimb}
\end{eqnarray}
Under this condition the oscillations in the matter of the Earth do not suppress the signal.  \\

\noindent 
(2)
The phase of oscillations at IMB at $E\simeq 50\div 60$ MeV is a semi-integer 
of $\pi$: 
\begin{eqnarray} 
\phi_{IMB}(60)= \pi \left({1 \over 2}+k \right)~,~~~~k=0,1,2,..~~,
\label{eq:phasimbsemi}
\end{eqnarray}
so that in this range of energy one expects maximal suppression effect.
%This also ensures that at $E\simeq 50$ MeV one expect maximal suppression of t%he signal due to oscillations in the Earth.  
It is easy to check that, in the range of parameters of interest, the 
conditions (\ref{eq:phasimb})  and (\ref{eq:phasimbsemi})  are satisfied simultaneously with good precision.
\\

\noindent 
(3) The phase of oscillations at K2 at $E\simeq 38\div 42$ MeV  is  
\begin{eqnarray} 
\phi_{K2}(40)\equiv {\pi d_{K2}\over l_m}=\pi \left({1 \over 2}+k \right)~,~~~~k=0,1,2,...~,
\label{eq:phask2}
\end{eqnarray}
so that the Earth matter effect produces maximal suppression of the K2 signal in the range  $E\simeq 38\div 42$ MeV.\\

\noindent 
(4) The Earth matter effect is maximal, $A_p\simeq A^{max}_p$, at IMB at the energies $E\simeq 50\div 60$ MeV, that is:
\begin{eqnarray} 
E^{IMB}_R \simeq 50\div 60~{\rm MeV}~.
\label{eq:maxsup}
\end{eqnarray}
\\

\begin{figure}[hbt]
\begin{center}
%\psfrag{MMMM}{$\Delta m^2 $}
%\psfrag{CCCCC}{${\cos 2\theta}$}
\epsfig{file=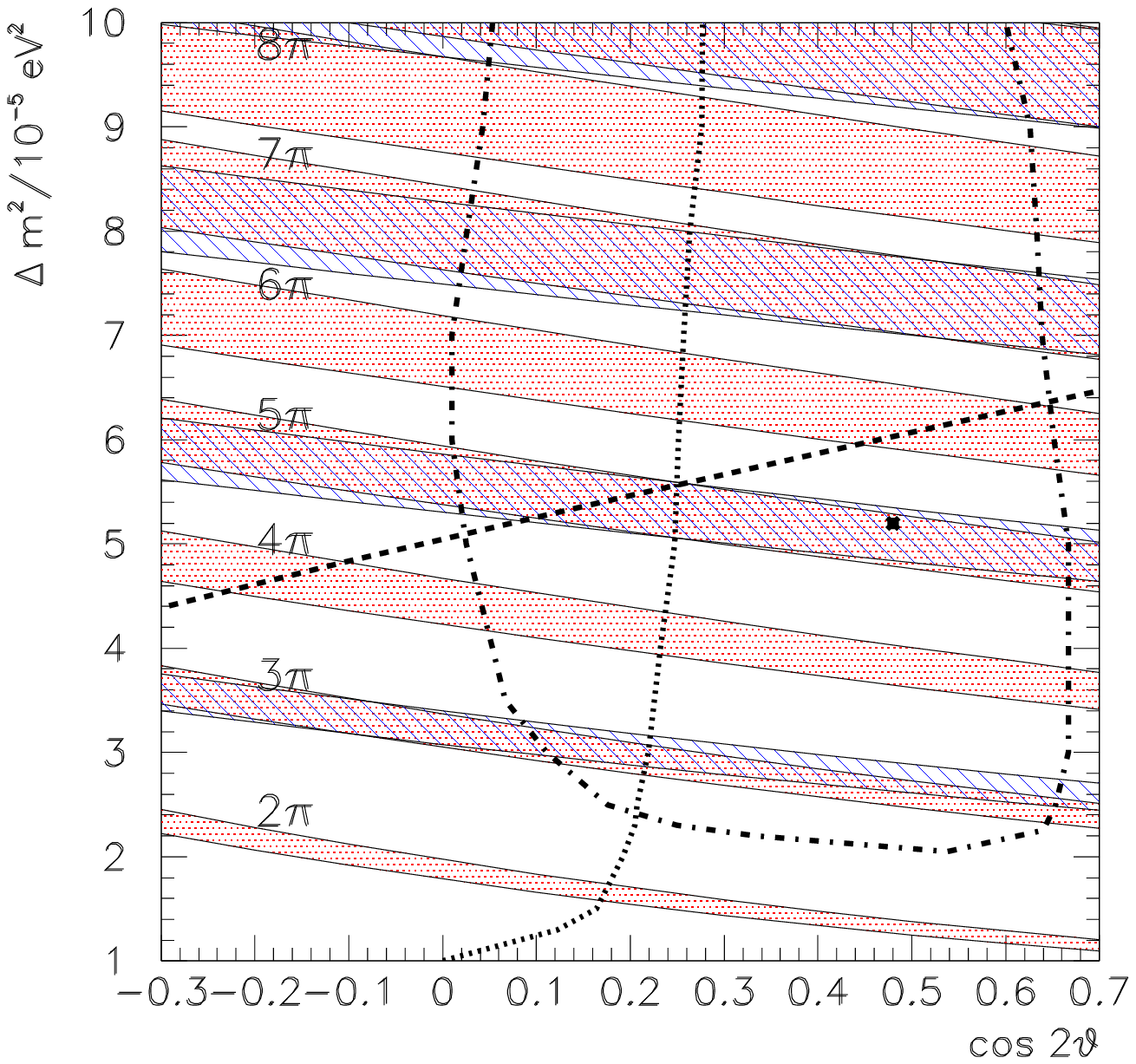, width=10truecm}
\end{center}
\caption{Bands of equal phases $\phi_{IMB}(40)= k \pi$ (dotted regions) and $\phi_{K2}(40)=\pi \left({1/2}+k \right)$ (dashed regions) in the $\Delta m^2$-$\cos 2\theta $ plane. The widths of the bands are determined by the requirement that the conditions (\ref{eq:phask2}) and (\ref{eq:phasimb}) are satisfied in the energy interval $E=38\div 42$ MeV.
The region below the dashed line represents the band of strong Earth matter effect, where 
$A_p\gta 0.7 A^{max}_p$ (see eq. (\ref{eq:maxsup})). 
For comparison we show the $99\%$ C.L. allowed region of the LMA solution of the solar neutrino deficit (dotted-dashed contour, from \protect\cite{Gonzalez-Garcia:2000sk}) where the best fit point is marked by an asterisk.  
The dotted line represents the $99\%$ C.L. exclusion curve from fig. 3a of ref. \protect\cite{Smirnov:1994ku}. 
} 
\label{fig:fig1} 
\end{figure}

In fig. \ref{fig:fig1} we show the conditions (\ref{eq:phasimb}), (\ref{eq:phask2}) and (\ref{eq:maxsup}) in the $\Delta m^2-\cos 2\theta$ plane.
 As follows from the figure, there are bands in which the requirements (\ref{eq:phasimb}) and (\ref{eq:phask2}) are satisfied simultaneously. They correspond to $\phi_{IMB}\simeq 2 \phi_{K2}=3\pi,5\pi,7\pi,...$~. The phase increases with $\Delta m^2$. Notice that the requirements (\ref{eq:phasimb})-(\ref{eq:phask2}) are satisfied in the whole  relevant range of $\cos 2\theta$ if  $\phi_{IMB}$  equals odd multiples of $\pi$.

The condition of maximal effect, eq. (\ref{eq:maxsup}), gives  
$\Delta m^2=(1.7\div 2.1)\cdot 10^{-5}~{\rm eV^2}$. Large Earth matter effect, e.g. $A_p\gta 0.7 A^{max}_p$, is realized in much wider interval, whose borders depend on $\cos 2\theta$ (the upper border is represented by the dashed line 
in fig. \ref{fig:fig1}).

As follows from fig. \ref{fig:fig1} the explanation of the properties (i)-(iii) implies the oscillation parameters to be in the regions  $\phi_{IMB}=3\pi$ for all the relevant values of $\theta$, and $\phi_{IMB}=5\pi$ for $\cos 2\theta\gta 0.2$.  The central values of these bands are described, approximatively, by the following lines:  
\begin{eqnarray} 
&&\Delta m^2\simeq 3.3 \cdot 10^{-5}~{\rm eV^2}\left[ 1-0.35\cos 2\theta\right]~, \nonumber \\
&&\Delta m^2\simeq 5.6 \cdot 10^{-5}~{\rm eV^2}\left[ 1-0.18 \cos 2\theta\right]~.
\label{eq:m2bands}
\end{eqnarray}
For $\cos 2\theta \gta 0.1$ the values (\ref{eq:m2bands}) are well inside the $99\%$ C.L. allowed region of the LMA solution (dotted-dashed contour in fig. \ref{fig:fig1}, from ref. \cite{Gonzalez-Garcia:2000sk}).  Moreover, the best-fit point, $\cos 2\theta\simeq 0.5$, $\Delta m^2\simeq 5.2 \cdot 10^{-5}~{\rm eV^2}$~\footnote{This result is given by the global two-neutrinos fit of the solar neutrino data including the total rates and the day and night spectra at Super-Kamiokande \cite{Gonzalez-Garcia:2000sk}.}, lies in the band $\phi_{IMB}\simeq 5\pi$ (see eq. (\ref{eq:m2bands})).   

The fig. \ref{fig:fig1} shows also the $99\%$ C.L. exclusion curve obtained in ref. \cite{Smirnov:1994ku} (dotted line), corresponding to the upper bound on the permutation parameter $p=0.35$. One can see that the region 
$\cos 2\theta \lta 0.25$ is excluded;  this region was obtained taking $T_{\bar{e}}\simeq 4.5$ MeV and $T_{\bar{\mu}'}\simeq 7.4$ MeV; it becomes narrower for the lower values of the temperatures we consider in this paper.

%For the solar neutrino best fit value, $\sin^2 2\theta\simeq 0.75$, 
%we find that $\Delta m^2$ should be in the intervals 
%$\Delta m^2=(2.4\div 2.8)\cdot 10^{-5}~{\rm eV^2}$ or 
%$\Delta m^2=(4.7\div 5.3)\cdot 10^{-5}~{\rm eV^2}$, which are well 
%inside the range of the preferable $\Delta m^2$  values of the LMA 
%solution region. 
 
The expected distributions of events are rather sensitive to variations of  $\Delta m^2$.
 For instance, a change of $\Delta m^2$ by $\sim 20\%$ in the 
 $\phi_{IMB}=3\pi$ band will lead to the prediction of strong suppression of the number of events in the interval $E\simeq 35\div 40$ MeV of the IMB spectrum and enhancement of the signal at  $E\simeq 50$ MeV, in contradiction with observations.
 
Let us consider the expected spectra of events at K2 and IMB. The original instantaneous $\bar{\nu}_e$ and $\bar{\nu}_{\mu'}$ fluxes can be described by ``pinched'' Fermi-Dirac spectra with pinching parameter $\eta \simeq 2$ \cite{Janka}. However, once integrated over arrival time intervals $\Delta t$ of several seconds, the spectra are well approximated by the ordinary Fermi-Dirac form, i.e. with $\eta=0$, as an effect of the decay with time of the neutrino luminosities and temperatures.  This description is good if $\Delta t$ is not larger than the typical decay time $\tau$ of the temperatures and luminosities.  For $\Delta t\gta \tau$ the integrated spectra can not be approximated by a Fermi-Dirac form.  This is the case of the SN1987A data, which show a rapid cooling of the neutrino spectra with decay time comparable or smaller than the duration of the burst,  $ \tau \lta \Delta t_b\simeq 13$ s.
Therefore, we have divided the whole time interval of observations in two bins, $t_1=0\div 6.5$ s and  $t_2=6.5\div 13$ s, and described the integrated fluxes over each bin by Fermi-Dirac spectra with different temperatures ($T(t_1)>T(t_2)$) and luminosities ($L(t_1)>L(t_2)$). 
We mark that the values of $T$ and $L$ we give should be considered as effective (time averaged) quantities.

\begin{figure}[hbt]
\begin{center}
\psfrag{E}{$E/{\rm MeV}$}
\psfrag{N}{${\rm N_{ev}}$}
\epsfig{file=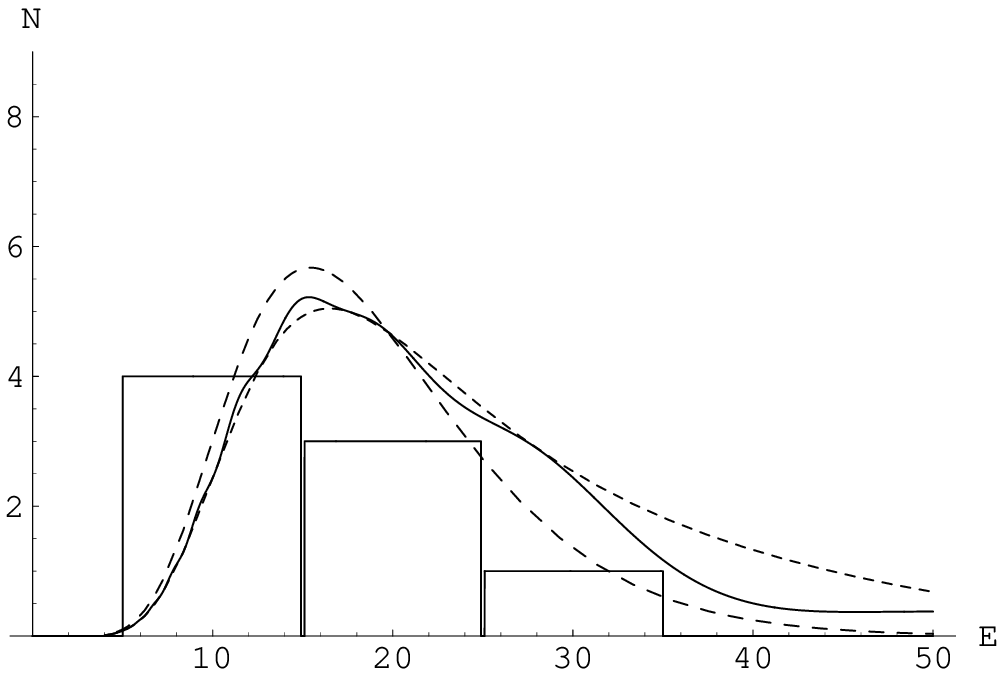, width=11truecm}
\epsfig{file=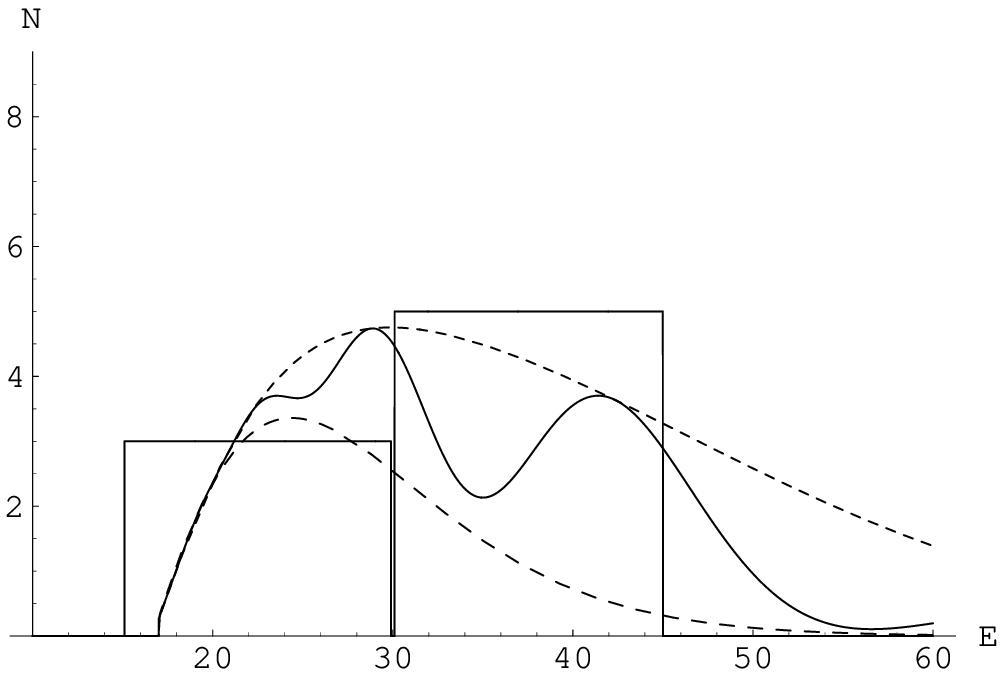, width=11truecm}
\end{center}
\caption{The predicted spectra of $\bar{\nu}_e$ events at Kamiokande-2 
(upper panel) and IMB (lower panel). We show: the original spectra without 
oscillation effects (long dashed lines), spectra with conversion in the star only (dashed lines), spectra with oscillation effects both in the star and in the Earth (solid lines).
We used the following set of oscillation parameters and characteristics of the original  $\bar{\nu}_e$ and $\bar{\nu}_\mu'$ spectra:  
$T_{\bar{e}}=3.5$ MeV, $L_{\bar{e}}=3 \cdot 10^{52}$ ergs, $\eta_{\bar{e}}=0$,  $T_{\bar{\mu}'}=7$ MeV, $L_{\bar{\mu}'}=3 \cdot 10^{52}$ ergs, $\eta_{\bar{\mu}'}=0$, $\cos 2\theta=0.5$, $\Delta m^2=2.75\cdot 10^{-5}~{\rm eV^2}$. 
The histograms show the observed distributions of events during the first 6.5 seconds.} 
\label{fig:fig2} 
\end{figure}

\begin{figure}[hbt]
\begin{center}
\psfrag{E}{$E/{\rm MeV}$}
\psfrag{N}{${\rm N_{ev}}$}
\epsfig{file=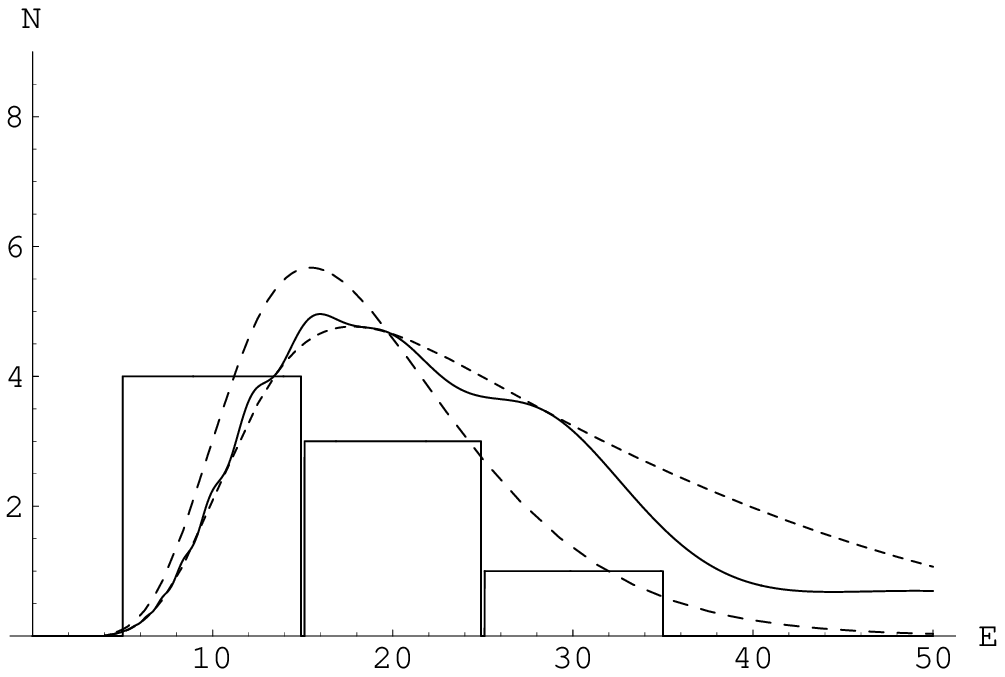, width=11truecm}
\epsfig{file=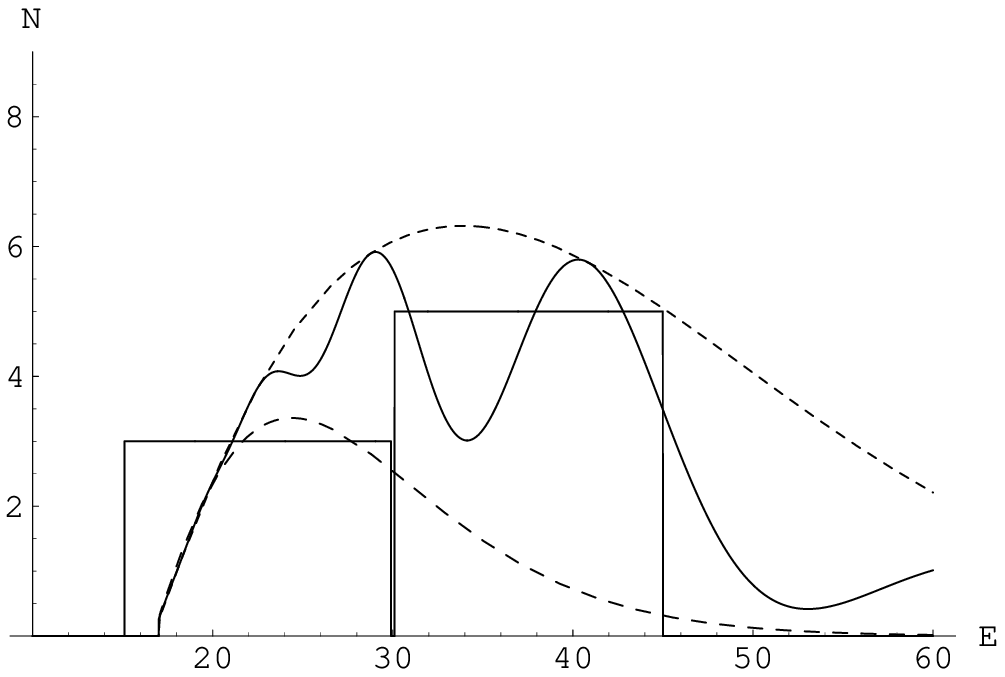, width=11truecm}
\end{center}
\caption{The same as fig. \ref{fig:fig2} with 
 $\cos 2\theta=0.2$, $\Delta m^2=3\cdot 10^{-5}~{\rm eV^2}$. } 
\label{fig:fig3} 
\end{figure}

In figures \ref{fig:fig2}-\ref{fig:fig3} we show the expected spectra of events at K2 and IMB in the first time bin ($t<6.5$ s) for two sets of parameters from the preferable regions (\ref{eq:m2bands}). 
They have been obtained taking  $T_{\bar{e}}=3.5$ MeV,   $T_{\bar{\mu}'}=7$ MeV and $\eta_{\bar{\mu}'}=\eta_{\bar{e}}=0$. 
We assumed equal integrated luminosities in  $\bar{\nu}_e$ and $\bar{\nu}_{\mu'}$:  $L_{\bar{e}}=L_{\bar{\mu}'}=3\cdot 10^{52}$ ergs.
The energy thresholds and detection efficiencies have been taken into account.

According to the figures, the predicted spectra with oscillations (solid 
lines) fit better the observed distributions of events at K2 and IMB. 
As expected, the spectra without oscillations (dashed lines) can not describe the concentration of IMB events at $E\simeq 30\div 40$ MeV and the absence of an excess of K2 events in this range.   The conversion in the star only leads to an appearance of high energy tails (short dashed lines) which contradict the observations of both K2 and IMB.  The oscillations in the matter of the Earth (solid lines) suppress 
the tails above $45$ MeV in IMB and above  $35$ MeV in K2~\footnote{Notice that in our analysis the maximal suppression, due to oscillations in the Earth, of the antineutrino signal in IMB occurs at energies above 50 MeV.  For such energies the efficiency of the IMB detector was high (${\cal E}_{IMB}\gta 0.7$).}.  Notice that the figures have illustrative character and do not correspond to the optimal (best fitted) choice of parameters of the original spectra. \\

To quantify the improvement of the fit due to conversion inside the star and  
the Earth matter effect we have performed the likelihood analysis following, 
in general, the prescriptions of ref. \cite{Jegerlehner:1996kx}.  There are, however, some differences:\\

\noindent
1). We have divided the whole energy range of the detected events into several bins. The size $\Delta E$ of each bin has been chosen according to the experimental errors $\epsilon$ in the measurements of energy: $\Delta E\sim 2\epsilon$.  We used three bins with $\Delta E= 10$ MeV for K2 and two bins with  $\Delta E=15$ MeV for IMB (see histograms in the figs. \ref{fig:fig2}-\ref{fig:fig3}). \\

\noindent
2). We computed the realization probability $P$ of a given theoretical prediction using the Poissonian distribution:
\begin{eqnarray} 
P=\prod_{i} {(n_i)^{N_i}\over N_i !} e^{-n_i}~,
\label{eq:poisson}
\end{eqnarray}   
where $ N_i$ is the number of observed events in the $i$-th bin and 
$n_i=n_i(T_{\bar{e}},T_{\bar{\mu}'},L_{\bar{e}},L_{\bar{\mu}'},$ $\Delta m^2,$ $\theta)$ is the corresponding number of expected events which depends on the parameters of the original neutrino spectra and oscillation parameters.
We find the maximum of $P$ and the contours of given confidence levels according to ref. \cite{Jegerlehner:1996kx}.  

Notice that, in contrast with \cite{Jegerlehner:1996kx} we use a discrete (binned) expression of the probability $P$ and we calculate $n_i$ by integrating the corresponding predicted function $n(E)$ over the  $i$-th energy bin. By this procedure we take into account the experimental errors on energies with no need to introduce any energy resolution function as it is done in \cite{Jegerlehner:1996kx}. \\

\noindent
3). We performed separate analyses of the K2 and IMB data collected in the two time bins $t_1$ and $t_2$; no events are discarded in this procedure\footnote{The event number 6 in K2 has been considered as due to background, following ref. \cite{Jegerlehner:1996kx}; however our conclusions remain unchanged when this event is considered.}. \\

\begin{figure}[hbt]
\begin{center}
\psfrag{T}{${T/{\rm MeV}}$}
\psfrag{L}{${L/{10^{53}~{\rm ergs}}}$}
\epsfig{file=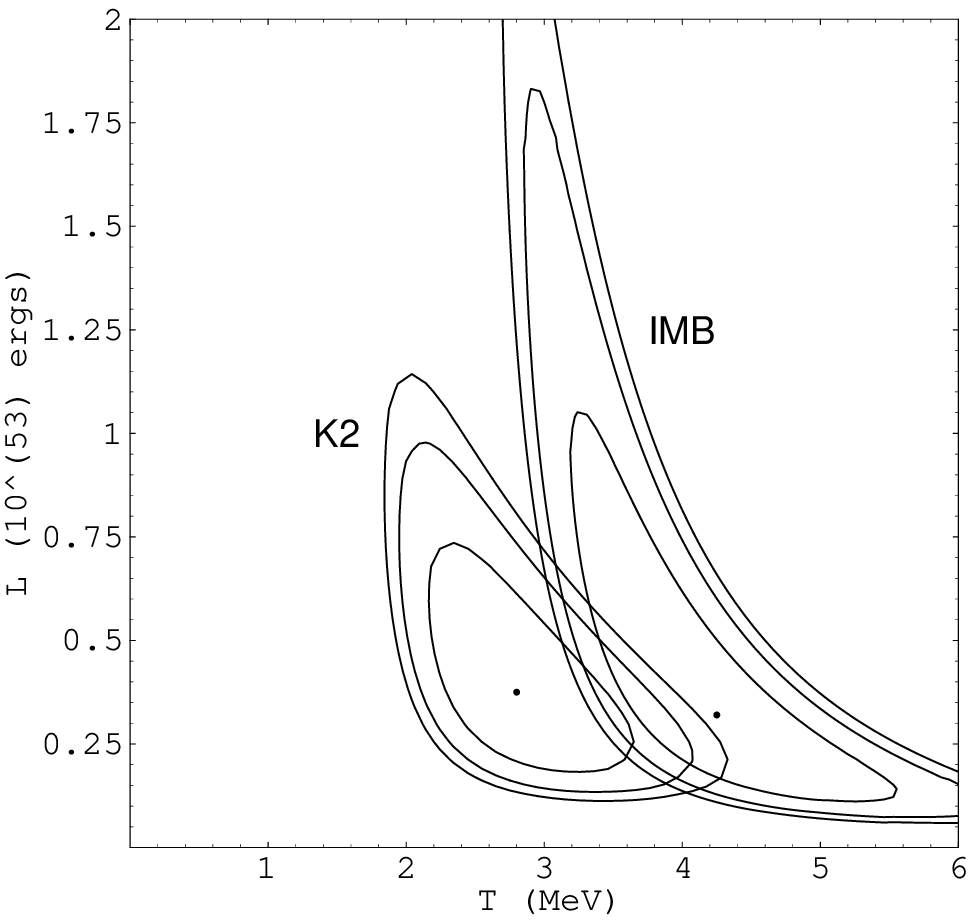, width=10truecm}
\epsfig{file=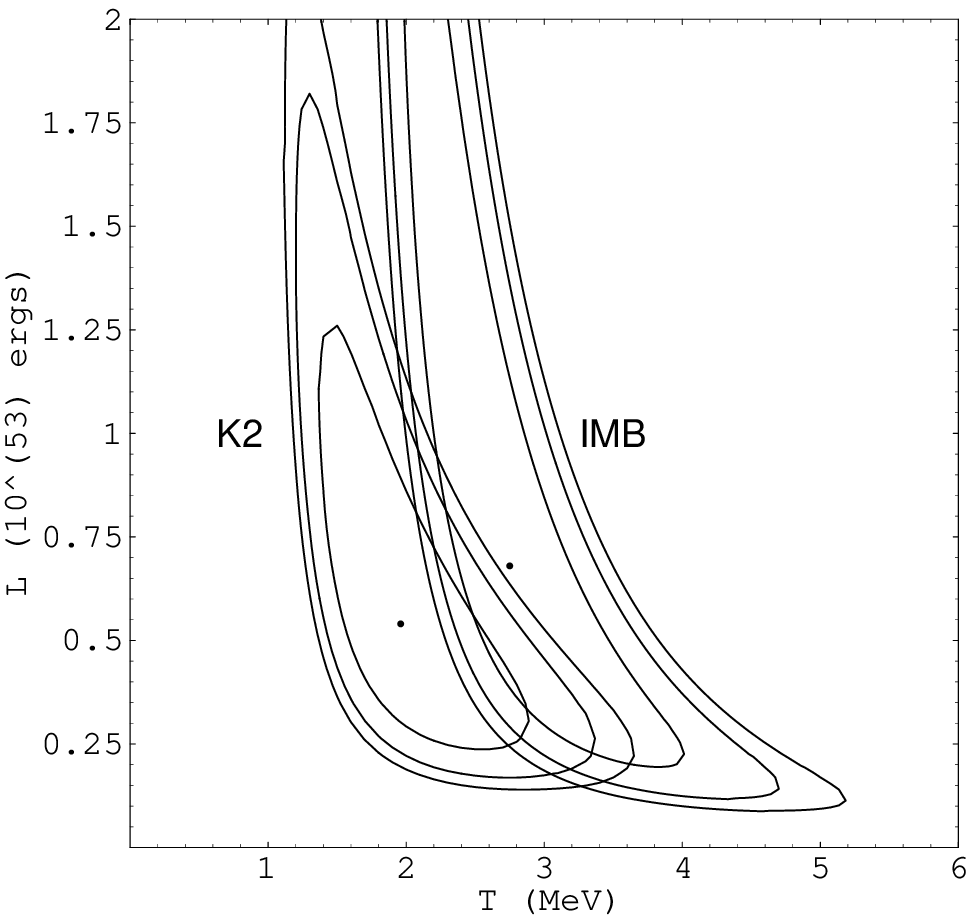, width=10truecm}
\end{center}
\caption{Best fit points and contours of equal $68,90,95.4\%$ likelihood for K2 and IMB in the $T_{\bar{e}}-L_{\bar{e}}$ plane. The upper panel shows the result of the separate fits of K2 and IMB data without oscillation effects. The lower panel represents a similar fit in presence of oscillations. The following values for the oscillation parameters and characteristics of original spectra have been used: 
$T_{\bar{\mu}'}/T_{\bar{e}}=1.8$,  $L_{\bar{\mu}'}/L_{\bar{e}}=1$,  $\eta_{\bar{e}}=\eta_{\bar{\mu}'}=0$ and $\cos 2\theta=0.2$, $\Delta m^2=3\cdot 10^{-5}~{\rm eV^2}$.
} 
\label{fig:fig4} 
\end{figure}

\noindent
In fig. \ref{fig:fig4} we show  the results of the likelihood analysis of the 
IMB and K2 data of the first time bin in the  $T_{\bar{e}}- L_{\bar{e}}$ plane in absence of mixing, $\theta=0$ (upper panel).
As follows from the figure, K2 and IMB imply substantially different temperatures of the original $\bar{\nu}_e$ spectrum: $T_{K2}\simeq 2.8 $ MeV and $T_{IMB}\simeq 4.2$ MeV. There is no overlap of the $68\%$ C.L. regions.

The lower panel of fig. \ref{fig:fig4} shows the analogous plot in presence of  oscillations in the star and in the matter of the Earth. We have taken  $T_{\bar{\mu}'}/T_{\bar{e}}=1.8$,  $L_{\bar{\mu}'}/L_{\bar{e}}=1$,  $\eta_{\bar{e}}=\eta_{\bar{\mu}'}=0$ with oscillation parameters from the preferable bands (see fig. \ref{fig:fig1}).
With oscillations the best fit parameters inferred from IMB and K2 become closer: $T_{K2}\simeq 1.96 $ MeV and $T_{IMB}\simeq 2.75$ MeV. 
We get $\Delta T/T\equiv 2(T_{IMB}-T_{K2})/(T_{IMB}+T_{K2})\simeq 0.33$ and 0.41 with and without oscillations respectively.
Overlap of the  $68\%$ C.L. regions appears. 
The combined fit gives $T_{\bar{e}}\simeq 2.77 $ MeV and $L_{\bar{e}}\simeq 4.4\cdot 10^{52}$ ergs, which are 
in good agreement with recent calculations \cite{Burrows:1999es}.
Thus, the likelihood analysis with oscillations shows some improvement of the fit and does not imply too low temperatures and high luminosities.

In the second time bin K2 has detected 3 events with low energies, close to the threshold, whereas IMB has no signals.  This can be easily interpreted in the assumption of lower temperatures of the $\bar{\nu}_e$ and $\bar{\nu}_{\mu'}$ original spectra and smaller difference of  $T_{\bar{\mu}'}$ and  $T_{\bar{e}}$. 
Taking for instance 
$T_{\bar{e}}\simeq T_{\bar{\mu}'}\simeq 1.8 $ MeV and $L_{\bar{e}}\simeq L_{\bar{\mu}'}\simeq 3\cdot 10^{52}$ ergs we predict $\sim 2$ events in K2 with energy $E\simeq 5\div 15$ MeV and less than 0.5 events at $E\geq 15$ MeV.
With the same values of the parameters we get less than 0.2 events in IMB, that is, with high probability all the events are below the IMB threshold. \\

For neutrino parameters from the LOW solution: $\Delta m^2=(0.3\div 2)\cdot 10^{-7}~{\rm eV^2}$, $\sin^2 2\theta\gta 0.95$,  we get $l_0/l_\nu\lta 10^{-2}$, 
so that the mixing angle in matter  is suppressed: $\sin 2\theta_m\simeq \sin 2\theta (l_0/l_\nu)$.   For the depth of oscillations in the Earth matter we find from eq. (\ref{eq:ap}):
\begin{eqnarray} 
A_p\simeq \sin^2 2\theta \left({l_0 \over l_\nu} \right)~,
\label{eq:alow}
\end{eqnarray}
that is, $A_p\ll 1$. Thus, the effect of oscillations in the Earth is negligibly small.  Now $P_{1 \bar{e}}\simeq \cos^2 \theta$ and the $\bar{\nu}_e$ flux at Earth is determined by the conversion inside the star only (see eq. (\ref{eq:flux}) and the short dashed lines in the figures \ref{fig:fig2}-\ref{fig:fig3}).  No improvement of the fit can be obtained in this case.  

\section{Implications for the neutrino mass spectrum}
\label{sect:4}
The fact of observation of the Earth  matter effect  in the $\bar{\nu}_e$ channel by itself has important implications for the neutrino mass spectrum and mixing (see also \cite{Dighe:2000bi}). \\

\noindent
1). As we have shown, a significant effect is possible for oscillation parameters in the LMA region only: thus, the observation of the effect selects the LMA solution of the solar neutrino problem.  In the cases of SMA or LOW solutions the Earth matter effect would be practically unobservable. \\

\noindent
2). Important conclusions can be drawn on the type of mass hierarchy and the mixing element $U_{e 3}$ in the three neutrino context.  Let us consider the three neutrino scheme which explains also the atmospheric neutrino problem via $\nu_\mu \leftrightarrow \nu_\tau$ oscillations.  In this scheme the third mass eigenstate, $\nu_3$, is isolated from $\nu_1$ and $\nu_2$, which are  responsible for the solar neutrino oscillations, by the mass gap $\Delta m^2_{2 3}\simeq \Delta m^2_{atm}\simeq 3\cdot 10^{-3}~{\rm eV^2}$.  The state $\nu_3$  has nearly maximal mixture of $\nu_\mu $  and 
$\nu_\tau$.  The admixture of the $\nu_e$ in this state, described by the matrix element  $|U_{e 3}|^2$, is small being restricted by the reactor experiments CHOOZ \cite{chooz} and  Palo Verde \cite{paloverde}: $|U_{e 3}|^2\lta 0.02\div 0.05$.

Observation of the matter effect means that (i) either the mass hierarchy is normal: $\Delta m^2_{2 3}=m^2_3-m^2_2>0$ (the third state is the heaviest one) or, (ii)  if the hierarchy is inverted,  $\Delta m^2_{2 3}<0$, the $\nu_e-\nu_3$ mixing is very small: 
$|U_{e 3}|^2\ll 10^{-3}$.
  
Indeed, in the first case the high density resonance associated with $\Delta m^2_{2 3}$ ($\rho_R\sim \Delta m^2_{2 3} $) is in the neutrino channel. It does not influence the conversion of antineutrinos and the situation coincides with the one considered above.  

If, however, the hierarchy is inverted the high density resonance is in the antineutrino channel. The pattern of conversion then strongly depends on the adiabaticity in this resonance.  If $|U_{e 3}|^2\gta 10^{-3}$ the level crossing is adiabatic, so that in the star the conversions 
\begin{eqnarray} 
\bar{\nu}_e \rightarrow \bar{\nu}_\tau'~,~~~  \bar{\nu}_\tau' \rightarrow \bar{\nu}_\mu'/\bar{\nu}_e~,~~{\rm and}~~  
\bar{\nu}_\mu' \rightarrow \bar{\nu}_\mu'/\bar{\nu}_e 
\label{eq:convocc}
\end{eqnarray}
occur. Thus, the $\bar{\nu}_e$ flux at Earth will be composed of converted $\bar{\nu}_\mu'$ and $\bar{\nu}_\tau'$ fluxes. Since the original $\bar{\nu}_\mu'$ and $\bar{\nu}_\tau'$  fluxes are identical, no matter effect should  be seen in the $\bar{\nu}_e$ flux \cite{Dighe:2000bi,talkminakata}. 
Moreover, the $\bar{\nu}_e$ flux will have the hard spectrum of the original 
$\bar{\nu}_\mu'$ and $\bar{\nu}_\tau'$  fluxes, which is disfavoured by observations \cite{Smirnov:1994ku}.

If $|U_{e 3}|^2\ll 10^{-3}$,  the adiabaticity in the high density resonance is broken. The effect of this resonance can be neglected and the pattern of the 
three neutrinos conversion is reduced to the two neutrino case discussed in the paper.

If $|U_{e 3}|^2\simeq  10^{-5}\div 10^{-3}$, one can observe an intermediate situation: harder $\bar{\nu}_e$ spectrum and partially suppressed Earth matter effect.  

If $|U_{e 3}|^2\gta 10^{-3}$ (which looks rather plausible in view of the large mixings between the first and the second as well as the second and the third generations) we come to the conclusion that the observation of the Earth matter effect in the   $\bar{\nu}_e$ channel implies a scheme with normal mass hierarchy and the LMA solution of the solar neutrino problem. \\

\noindent
3). Let us consider the Earth matter effect for supernova neutrinos in the case of $\nu_e-\nu_s$ mixing. As for active neutrinos, for $\nu_e-\nu_s$  a significant effect is possible for large mixing angles (notice, however, that large mixing $\nu_e\rightarrow \nu_s$ conversion gives poor fit of the solar neutrino data).

We assume that no sterile antineutrino fluxes are generated in the central part of the star\footnote{A sterile state $\nu_s$ can  be produced, however, by conversion of active neutrinos at high densities in the star. This happens in some schemes with four neutrinos which explain the LSND result.}. In this case we get the result for the $\bar{\nu}_e$ flux at the detector from eq. (\ref{eq:combin}) putting $F^0_{\bar{\mu}'}=0$:
\begin{eqnarray} 
F^D_{\bar{e}}=P^{s}_{1 \bar{e}} F^0_{\bar{e}}~.
\label{eq:nus}
\end{eqnarray}
Here $P^{s}_{1 \bar{e}}$ is the probability of $\bar{\nu}_1\rightarrow \bar{\nu}_e$ oscillations in the matter of the Earth for mixing with sterile neutrino.

The difference of potentials for the $\nu_e-\nu_s$ system, $V_{e s}=\sqrt{2}G_F\left(n_e -n_n/2\right)\simeq V_{e \mu}/2$, is about two times smaller than 
the one of the $\nu_e$ and $\nu_\mu$ species, 
$V_{e \mu}$. 
In the last equality we considered that the medium is almost isotopically neutral in the mantle of the Earth.  Correspondingly, the refraction length is two times larger than in the active-active case, so that,  according to eq. (\ref{eq:ll0}) the range of large matter effects is shifted to smaller $\Delta m^2$ by a factor 2.

From eq. (\ref{eq:nus}) we see that, in contrast with $\nu_e-\nu_\mu'$ mixing, the observed events in the detector are due to the original $\bar{\nu}_e$ flux only, and no hard tail appears in the spectrum.  As a consequence, no improvement in the fit of K2 and IMB data is obtained with respect to the $\theta=0$ case. Moreover, it is possible to check that a larger $\bar{\nu}_e$ luminosity is required.

The discussion of the sections \ref{sect:2}-\ref{sect:3} can be immediately generalized to the case in which the electron neutrino is mixed with both an active neutrino $\nu_\mu$ and a sterile one, $\nu_s$. The problem can be reduced to $\nu_e-\nu_x$ mixing, with $\nu_x=\sqrt{1-\delta}\nu_\mu + \sqrt{\delta}\nu_s$; the pure active and pure sterile mixings correspond to $\delta=0$ and  $\delta=1$ respectively.   
Now the $\bar{\nu}_e$ flux in the detector is given by eq. (\ref{eq:combin}) with $F^0_{\bar{x}}=(1-\delta) F^0_{\bar{\mu}'}$. 
Thus, with respect to the pure active case, one gets a reduction by a factor 
$\sim(1-\delta)$ of the hard part of the detected spectrum. For $\delta \lta 0.5$ still one gets a good fit of the K2 and IMB signals.

\section{Conclusions}
\label{sect:concl}
We have studied properties of the oscillation effects in the matter of the Earth on neutrino signals from supernovae.  We show that certain features of the
 neutrino signal from SN1987A detected by K2 and IMB detectors (difference of spectra, absence of events above $E\sim 40$ MeV) can be explained by oscillations in the Earth and different positions of the detectors. 
The consistency of the $\bar{\nu}_e$ spectra implied by K2 and IMB data improves when oscillations are taken into account.

The existence of the Earth matter effects on the neutrino burst from supernovae has crucial implications for the neutrinos mass spectrum:

\noindent
(1) The oscillation parameters $\Delta m^2$  and $\sin^2 2\theta$ should be in the range of LMA solution of the solar neutrino problem. Moreover, the values of 
 $\Delta m^2$  and $\sin^2 2\theta$ should lie in rather narrow bands.

\noindent
(2) The mass hierarchy (in the three neutrino context) should be normal, if 
 $|U_{e 3}|^2\gta 10^{-3}$ and it can be inverted provided that  $|U_{e 3}|^2\ll 10^{-3}$.

It should be stressed that,
in view of the low statistics of SN1987A neutrino signals, our conclusions have an indicative character, as far as the interpretation of the K2 and IMB data is concerned.
If the LMA solution will be identified in future experiments, 
 the inclusion of the effects of oscillations in the matter of the Earth in the analysis of SN1987A data will be unavoidable, and 
 the interpretation of SN1987A signals given in this paper will be confirmed.
The analysis performed in this paper shows that searches for oscillation effects in the matter of the Earth in future observations  of galactic supernovae will give important information on the neutrino mass spectrum, independently of supernova models.

\section*{Acknowledgments}
The authors wish to thank H.~Minakata, E.~Nardi, H.~Nunokawa, C.~Pe\~na-Garay and O.L.G.~Peres for valuable discussions and useful suggestions.

\subsection*{Note added}
In the recent paper \cite{Cline:2000gu} a new analysis of the SN1987A signal is performed, based on the Kolmogorov test of the K2 data only, 
with the conclusion that the LMA solution of the solar neutrino problem is excluded at high confidence level.
However the Earth matter effect, which we have shown here to be crucial, is not taken into account.

%\begin{figure}[hbt]
%\begin{center}
%\psfrag{E}{$E/{\rm MeV}$}
%\psfrag{N}{${\rm N_{ev}}$}
%\epsfig{file=fig2b.eps, width=14truecm}
%\end{center}
%\caption{The same as fig. \ref{fig:fig2a} for IMB. } 
%\label{fig:fig2b} 
%\end{figure}

%\begin{figure}[hbt]
%\begin{center}
%\psfrag{E}{$E/{\rm MeV}$}
%\psfrag{N}{${\rm N_{ev}}$}
%\epsfig{file=fig3b.eps, width=14truecm}
%\end{center}
%\caption{The same as fig. \ref{fig:fig3a} for IMB. } 
%\label{fig:fig3b} 
%\end{figure}

%%%%%%%%%%%%%%%%%%%%%%%%%%%%%%%%%%%%%%%%%%%%%%%%%%%%%%%%%%%%%%%%%%%%%%%%%
%%%%%%%%%%%%%%%%%%%%%%%%%%%%%%%%%%%%%%%%%%%%%%%%%%%%%%%%%%%%%%%%%%%%%%%%%
\bibliography{sn}
\end{document}